\documentclass[12pt]{article}

\pdfoutput=1

\usepackage{cite}

\usepackage{amssymb,latexsym}

\usepackage{epstopdf}

\usepackage{graphics,epsfig}

\usepackage{color}

\usepackage{mathrsfs}

\DeclareMathAlphabet{\mathpzc}{OT1}{pzc}{m}{it}

\usepackage{tikz}
\usetikzlibrary{arrows,shapes}
\usetikzlibrary{trees}
\usetikzlibrary{matrix,arrows} 				
\usetikzlibrary{positioning}				
\usetikzlibrary{calc,through}				
\usetikzlibrary{decorations.pathreplacing}  
\usepackage{pgffor}							

\usetikzlibrary{decorations.pathmorphing}	
\usetikzlibrary{decorations.markings}
\tikzset{
    vector/.style={decorate, decoration={snake}, draw},
	provector/.style={decorate, decoration={snake,amplitude=2.5pt}, draw},
	antivector/.style={decorate, decoration={snake,amplitude=-2.5pt}, draw},
    fermion/.style={draw=black, postaction={decorate},
        decoration={markings,mark=at position .55 with {\arrow[draw=black]{>}}}},
    fermionbar/.style={draw=black, postaction={decorate},
        decoration={markings,mark=at position .55 with {\arrow[draw=black]{<}}}},
    fermionnoarrow/.style={draw=black},
    gluon/.style={decorate, draw=black,
        decoration={coil,amplitude=4pt, segment length=5pt}},
    scalar/.style={dashed,draw=black, postaction={decorate},
        decoration={markings,mark=at position .55 with {\arrow[draw=black]{>}}}},
    scalarbar/.style={dashed,draw=black, postaction={decorate},
        decoration={markings,mark=at position .55 with {\arrow[draw=black]{<}}}},
    scalarnoarrow/.style={dashed,draw=black},
    electron/.style={draw=black, postaction={decorate},
        decoration={markings,mark=at position .55 with {\arrow[draw=black]{>}}}},
	bigvector/.style={decorate, decoration={snake,amplitude=4pt}, draw},
}

\tikzstyle{block} = [draw, rectangle, 
    minimum height=3em, minimum width=6em]

\let\a=\alpha \let\b=\beta \let\g=\gamma \let\d=\delta \let\e=\epsilon
\let\z=\zeta  \let\th=\theta  \let\k=\kappa
\let\l=\lambda \let\m=\mu \let\n=\nu \let\x=\xi \let\p=\pi 
\let\s=\sigma   \let\f=\phi  
 
      \let\G=\Gamma  \let\Th=\Theta \let\L=\Lambda
\let\X=\Xi  \let\S=\Sigma  \let\Y=\Psi
 
\let\la=\label  
  
\def\nn{\nonumber} \def\bd{\begin{document}} \def\ed{\end{document}}
\def\ds{\documentstyle} \let\fr=\frac \let\bl=\bigl \let\br=\bigr
\let\Br=\Bigr \let\Bl=\Bigl
\let\bm=\bibitem
\let\na=\nabla
\def\tU{{\widetilde U}}
\let\pa=\partial \let\ov=\overline
\def\ie{{\it i.e.\ }}
\newcommand{\be}{\begin{equation}}
\newcommand{\ee}{\end{equation}}
\def\ba{\begin{array}}
\def\ea{\end{array}}
\def\ft#1#2{{\textstyle{{\scriptstyle #1}\over {\scriptstyle #2}}}}
\def\fft#1#2{{#1 \over #2}}
\def\F#1#2{{ F_{#1}^{(#2)} }}
\def\cF#1#2{{ {\cal F}_{#1}^{(#2)} }}

\def\R{{\bf R}}
\def\sst#1{{\scriptscriptstyle #1}}
\def\oneone{\rlap 1\mkern4mu{\rm l}}
\def\e7{E_{7(+7)}}
\def\td{\tilde}
\def\wtd{\widetilde}
\def\im{{\rm i}}
\def\bog{Bogomol'nyi\ }
\newcommand{\ho}[1]{$\, ^{#1}$}
\newcommand{\hoch}[1]{$\, ^{#1}$}
\newcommand{\bea}{\begin{eqnarray}}
\newcommand{\eea}{\end{eqnarray}}
\newcommand{\ra}{\rightarrow}
\newcommand{\lra}{\longrightarrow}
\newcommand{\Lra}{\Leftrightarrow}
\newcommand{\ap}{\alpha^\prime}
\newcommand{\bp}{\tilde \beta^\prime}
\newcommand{\cB}{{\cal B}}
\newcommand{\cO}{{\cal O}}
\newcommand{\vecx}{\vec{x}}
\newcommand{\vecy}{\vec{y}}
\newcommand{\vecp}{\vec{p}}
\newcommand{\vecq}{\vec{q}}
\newcommand{\tr}{{\rm tr} }
\newcommand{\Tr}{{\rm Tr} }
\newcommand{\NP}{Nucl. Phys. }

\newcommand{\cL}{{\cal L}}
\newcommand{\cA}{{\cal A}}
\newcommand{\cT}{{\cal T}}
\newcommand{\cR}{{\cal R}}
\newcommand{\cD}{{\cal D}}
\newcommand{\cH}{{\cal H}}

\def\Cb{\bar{C}}

\def\sst#1{{\scriptscriptstyle #1}}
\def\0{{\sst{(0)}}}
\def\1{{\sst{(1)}}}
\def\2{{\sst{(2)}}}
\def\3{{\sst{(3)}}}
\def\4{{\sst{(4)}}}
\def\5{{\sst{(5)}}}
\def\6{{\sst{(6)}}}
\def\7{{\sst{(7)}}}
\def\8{{\sst{(8)}}}
\def\9{{\sst{(9)}}}
\def\p{{\sst{(p)}}}
\def\q{{\sst{(q)}}}
\def\ve{\varepsilon}
\def\vf{\varphi}
\def\F{\Phi}
\def\wg{\wedge}

\def\thb{\bar{\theta}}
\def\Thb{\bar{\Theta}}
\def\barp{\bar{p}}
\def\barq{\bar{q}}
\def\barc{\bar{c}}
\def\bard{\bar{d}}
\def\e{\epsilon}

\def \bi{\bibitem}
\def \la {\label}

\def \l {\lambda}
\def\foot{\footnote}
\def \tl  {{\tilde \l}}
\def \sql {{\sqrt \l}}
\def \adss {$AdS_5 \times S^5$\ }
\newcommand{\rf}[1]{(\ref{#1})}
\def \ov {\over}

\def\th{\theta}
\def\Th{\Theta}
\def\vth{\vartheta}
\def\btheta{{\bar\theta}}
\def\ttheta{{{\tilde\theta}}}
\def\bttheta{{{\bar\ttheta}}}
\def\vth{\vartheta}

\def\ra{\rightarrow}
\def\N{\nabla}
\def\F{{\cal F}}
\def\uM{\underline{M}}
\def\uA{\underline{A}}
\def\uN{\underline{N}}
\def\uP{\underline{P}}
\def\ua{\underline{a}}
\def\ub{\underline{b}}
\def\uc{\underline{c}}
\def\ud{\underline{d}}
\def\ue{\underline{e}}
\def\uf{\underline{f}}
\def\ui{\underline{i}}
\def\uj{\underline{j}}
\def\uk{\underline{k}}
\def\ul{\underline{l}}
\def\ual{\underline{\alpha}}
\def\ube{\underline{\beta}}
\def\um{\underline{m}}
\def\un{\underline{n}}
\def\up{\underline{p}}
\def\uq{\underline{q}}
\def\ur{\underline{r}}
\def\us{\underline{s}}
\def\umu{\underline{\mu}}
\def\unu{\underline{\nu}}
\def\ula{\underline{\l}}
\def\uka{\underline{\k}}
\def\usi{\underline{\s}}
\def\urh{\underline{\r}}
\def\cc{\circ}
\def\eqv{\equiv}

\def\ni{\noindent}

\def\Ep{E^{{}^{(+)}}}
\def\Em{E^{{}^{(-)}}}

\def\Mp{M^{{}^{(+)}}}
\def\Mm{M^{{}^{(-)}}}

\def \ha{{1\ov 2}}

\def\r{\rho}

\def\Y{{\rm Y}}
\def\X{{\rm X}}
\def\tY{\tilde{\rm Y}}
\def\tX{\tilde{\rm X}}
\def\dY{\dot{\rm Y}}
\def\dX{\dot{\rm X}}

\def \J {\mathcal{J}}
\def \del {\partial}

\def\dF{\dot{F}}
\def\dG{\dot{G}}
\def\df{\dot{f}}
\def \E {{\cal E}}
\def \S {{\cal S}}
\def \J {{\cal J}}

\def\ms{\mathcal{S}}
\def\mj{\mathcal{J}}
\def\soj{\fr{\ms}{\mj}}
\def \R {{\bf R}}
\def \om {\omega}
\def \bE {\bar E}
\def \x {{\cal X}}

\def \bi{\bibitem}
\def \la {\label}

\def \l {\lambda}
\def\foot{\footnote}
\def \tl  {{\tilde \l}}
\def \sql {{\sqrt \l}}
\def \adss {$AdS_5 \times S^5$\ }
\def \ov {\over}

\def \varpi {{\rm w}}

\def\thb{\bar{\theta}}
\def\Thb{\bar{\Theta}}
\def\mb{\bar{\m}}
\def\ab{\bar{\a}}
\def\zb{\bar{z}}
\def\psib{\bar{\psi}}
\def\barp{\bar{p}}
\def\barq{\bar{q}}
\def\barc{\bar{c}}
\def\bard{\bar{d}}
\def\e{\epsilon}
\def\wb{\bar{w}}
\def\lb{\bar{\l}}
\def\Jb{\bar{J}}
\def\Nb{\bar{N}}
\def\Zb{\bar{Z}}
\def\pab{\bar{\pa}}

\def\At{\tilde{A}}
\def\Bt{\tilde{B}}
\def\Ct{\tilde{C}}
\def\Dt{\tilde{D}}
\def\Et{\tilde{E}}
\def\Ft{\tilde{F}}
\def\Gt{\tilde{G}}
\def\Ht{\tilde{H}}
\def\Kt{\tilde{K}}
\def\Mt{\tilde{M}}
\def\Nt{\tilde{N}}
\def\Rt{\tilde{R}}
\def\at{\tilde{a}}
\def\bt{\tilde{b}}
\def\ct{\tilde{c}}
\def\dt{\tilde{d}}
\def\et{\tilde{e}}
\def\ft{\tilde{f}}
\def\htil{\tilde{h}}
\def\gt{\tilde{g}}
\def\nt{\tilde{n}}
\def\mut{\tilde{\mu}}
\def\nut{\tilde{\nu}}
\def\pht{\tilde{\f}}
\def\Pht{\tilde{\Phi}}
\def\vft{\tilde{\vf}}
\def \zet{\tilde{\z}}

\def\rht{\tilde{\rho}}

\def\asth{\hat{*}}
\def\phh{\hat{\phi}}

\def\bA{{\bf A}}

\def\ola{\overleftarrow}
\def\ora{\overrightarrow}
\def\alt{\tilde{\a}}

\def\eh{\hat{e}}
\def\eph{\hat{\e}}
\def\ph{\hat{p}}
\def\alh{\hat{\a}}
\def\beh{\hat{\b}}
\def\gah{\hat{\g}}
\def\Fh{\hat{F}}
\def\muh{\hat{\m}}
\def\nuh{\hat{\n}}
\def\thh{\hat{\th}}
\def\rhh{\hat{\r}}
\def\dh{\hat{d}}
\def\ih{\hat{i}}
\def\jh{\hat{j}}
\def\hh{\hat{h}}
\def\nh{\hat{n}}
\def\gh{\hat{g}}
\def\kh{\hat{k}}
\def\deh{\hat{\d}}
\def\wh{\hat{w}}
\def\lah{\hat{\l}}
\def\Ah{\hat{A}}
\def\Kh{\hat{K}}
\def\Nh{\hat{N}}
\def\Rh{\hat{R}}
\def\Ch{\hat{C}}
\def\Omh{\hat{\Omega}}

\def\xh{\hat{x}}

\def\ps{\rlap{\, /}\;\,p }
\def\ks{\rlap{\, /}\;\,k }

\def\gym{g_{YM}}

\def\adot{\dot{a}}
\def\bdot{\dot{b}}
\def\bpa{\bar{\pa}}

\def\pr{\prime}
\def\ssk{\medskip}
\def\clb{\color{blue}}
\def\clr{\color{red}}
\def\clg{\color{green}}

\def\bfA{{\bf A}}
\def\bfB{{\bf B}}
\def\bfK{{\bf K}}
\def\bfU{{\bf U}}
\def\bfX{{\bf X}}
\def\bfY{{\bf Y}}
\def\bfZ{{\bf Z}}
\def\bfg{{\bf g}}
\def\bfn{{\bf n}}

\def \vk{\vec{k}}
\def \vx{\vec{x}}

\begin{document}

\overfullrule=0pt
\parskip=2pt
\parindent=12pt
\headheight=0in \headsep=0in \topmargin=0in
\oddsidemargin=0in

\vspace{ -3cm}
\thispagestyle{empty}

 \vspace{0.1cm}

\setcounter{equation}{0}
\setcounter{footnote}{0}
\setcounter{section}{0}

\begin{center}

{\Large\bf Quantum ``violation" of Dirichlet boundary condition}

\vskip 0.8cm

 \vspace{.5cm}

\vspace{0.5cm}
I. Y. Park
\\

\vspace{0.3cm}

\vspace{0.3cm}
{\it Department of Applied Mathematics,
Philander Smith College 
                               \\
Little Rock, AR 72202, USA \\
inyongpark05@gmail.com
}

\end{center}

 \vspace{0.1cm}

\begin{abstract}

Dirichlet boundary conditions have been widely used in general relativity. They seem at odds with the holographic property of gravity simply because a boundary configuration can be varying and dynamic instead of dying out as required by the conditions. In this work we report what should be a tension between the Dirichlet boundary conditions and quantum gravitational effects, and show that a quantum-corrected black hole solution of the 1PI action no longer obeys, in the naive manner one may expect, the Dirichlet boundary conditions imposed at the classical level.
We attribute the `violation' of the Dirichlet boundary conditions to a certain mechanism of the information storage on the boundary.

\end{abstract}
\newpage

\section{Introduction}

Quantum gravitational effects may hold the key to some of the outstanding problems in theoretical physics.
One of such problems may be the black hole information paradox \cite{Hawking:1976ra}.  In this work we analyze  the perturbative quantum effects on the boundary and Dirichlet boundary conditions imposed at the classical level. In particular we observe a tension between Dirichlet boundary condition and perturbative quantum corrections: for a black hole solution, the quantum-corrected solution no longer obeys the Dirichlet-type boundary condition in a naively expected manner.  
A certain mechanism of the information storage on the boundary seems responsible for the phenomenon invloved.

The study of the boundary degrees of freedom has a long history (see \cite{Smolin:1995vq,Carlip:1998wz,Modesto:2005sj,Oeckl:2003vu} to name just a few works). More recently there has been an interesting development in the context of loop quantum gravity \cite{Freidel:2016bxd,Donnelly:2016auv} wherein a new possibility regarding the boundary dynamics has been explored in a manner more concrete than ever: instead of subtracting out the boundary terms (arising from the bulk variation) by the Gibbons-Hawking term, one adds an additional boundary term in such a way to preserve the gauge symmetry of the bulk and boundary terms altogether.\footnote{It was stated that inclusion of all of the possible boundary conditions is necessary for the complete Hilbert space. This seems to be in line with the AdS/CFT, and the present work shares the spirit.} An explicit identification of the boundary dynamics has been made. In the present work, we will employ the conventional approach of subtracting the boundary term but will further comment on this new development.

The idea of the violation of the classical-level Dirichlet boundary condition pursued in the present work is rather simple and can be illustrated with a time- and position- dependent solution that satisfies the Dirichlet boundary conditions at the classical level but receives quantum corrections that do not necessarily preserve the condition. As we will explicitly demonstrate by taking a 4D gravity-scalar system with a negative cosmological constant, some of the surface terms arising from varying quantum correction terms would, unless subtracted out, lead to violation of the Dirichlet condition imposed on a classical black hole solution.

Although we believe that the phenomenon occurs generically, an explicit demonstration is highly technically complicated for several reasons.
Firstly, the demonstration requires consideration of a time- and position- dependent background.
In general, finding such a solution is a complicated problem even at the classical level and we will rely on a series form of the solution. 
Secondly, it is also necessary to compute the 1PI effective action in order to obtain the quantum-corrected equations of motion. For a simple background, one may manage to compute the 1PI action after the classical action is expanded around the background under consideration. This procedure becomes extremely complicated when the background is time- and position- dependent. It would be nontrivial to obtain even the Green's function, let alone the 1PI action.

Because these complications are cumbersome and inessential for observing the potential violation of the Dirichlet boundary condition, we take several simplifying steps. They are inessential because what we focus on are the finite pieces contained within the {\em divergences}. The divergent parts are common to any background since they are expected to arise from a flat limit.
To compute the 1PI action perturbatively it would be more appropriate to expand the action around the background of interest\footnote{It would not matter if all of the nonperturbative contributions could be included as well. Within the perturbative calculation and for more complete results including the genuinely finite parts (not just the finite piece contained in the divergent terms), however, the proper way should be to expand around the actual background under consideration.} and obtain the Green's function and  1PI action associated with that background. However, since this task is simply too complicated one may take a simpler but legitimate step: one may consider the vacuum, i.e., an AdS spacetime, expand the classical action around it and work out the 1PI action. Expanding around the AdS background cannot, in general, be entirely justified as a step to obtain the perturbative 1PI action (including the finite parts of the loop diagrams) from which the quantum effects on the boundary of another solution (namely, the time- and position- dependent solution) is studied. 
However, because we seek after the finite parts contained within the divergences ultimately, we can consider the AdS spacetime (and its flat space limit since the divergences must come from the flat limit). 
Therefore, the present procedure serves the purpose of revealing the quantum gravitational effects on the boundary condition. 
Throughout, we adopt the conventional framework (other than employing the traceless propagator \cite{Ortin,Park:2015ota,Park:2015xoa}) and limit the loop analysis to one-loop in dimensional regularization.

\section{Review}

We begin by reviewing the Dirichlet boundary condition and the time-dependent black hole solution \cite{Murata:2010dx}.
In section 3 we study how they are modified by the quantum effects.

\subsection{classical boundary condition}

As well known, the variation of the Einstein-Hilbert term yields the following boundary terms:
\bea
\int d^4x \sqrt{-g}\; \nabla^{\a}\Big[  \nabla^{\b}\d g_{\a\b}- \nabla_{\a} (g^{\k_1\k_2}\d g_{\k_1\k_2}) \Big] \la{veh}
\eea
Since the structure of these terms is such that the variations are acted on by covariant derivatives one cannot impose the Dirichlet condition.
For this reason one adds an extra term of the trace of the second fundamental form - often called the Gibbons-Hawking term \cite{Gibbons:1976ue} (see \cite{Wald,Poisson} for reviews; discussions on the Dirichlet boundary conditions can also be found, e.g., in \cite{Hartle:1981cf,Hayward:1993my,Mann:2005yr,Neiman:2012fx,Parattu:2015gga,Lehner:2016vdi,Krishnan:2016dgy}) - in the starting action so that the variation of the extra term exactly cancels the boundary terms above. It is then possible to impose the Dirichlet boundary condition. It would not be necessary to add the extra term and impose the Dirichlet boundary condition if the boundary term above could be made to vanish by itself through a different (say, Neumann-type) boundary condition. 
With the Gibbons-Hawking term added, the Dirichlet boundary condition is self-consistent at the classical level.

\subsection{classical time-dependent black hole solution}

The system we consider is the following gravity-scalar system:
\begin{equation}
S=\int d^3 x \sqrt{-g}\left[\fr1{\k^2}\Big(R+\frac{6}{L^2}\Big)-\fr12(\pa_\mu \z)^2 -\fr12m^2\z^2 \right] 
\end{equation}
where $m^2=-\fr{2}{L^2}$. We set the radius $L=1$ in some places and $\frac{6}{L^2}=-2 \L$ later. ($\L$ denotes the cosmological constant.) Consider the following metric ansatz \cite{Murata:2010dx}:
\begin{eqnarray}
ds^2&=&-\frac{1}{z^2}\left[F(t,z) dt^2 + 2 dt\,dz\right] + \Phi(t,z)^2 (dx^2+dy^2)\nonumber\\
\z&=&\z(t,z)\nonumber
\end{eqnarray}
with
\begin{eqnarray}
F(t,z)&=&F_0(t) +F_1(t) z+ F_2(t)z^2+F_3(t)z^3 + ...\nonumber\\
\Phi(t,z)&=&\frac{1}{z}+\Phi_0(t) +\Phi_1(t) z+ \Phi_2(t)z^2+\Phi_3(t)z^3 + ...\nonumber\\
\z(t,z)&=&\z_0(t) +\z_1(t) z+ \z_2(t)z^2+\z_3(t)z^3 + ... \la{sersol}
\end{eqnarray}
Substituting these ansatze into the field equations one gets:
\bea
&&\hspace{-.2in}\z _0=\Phi_0=F_1=0, \quad F_0=1, \quad  \Phi _1=-\frac{1}{8} \z _1^2,\quad F_2=-\frac{1}{4} \z _1^2,\quad
\Phi _2=-\frac{1}{6} \z _1 \z_2 \nn\\
&&\hspace{-.7in} \z _3=\frac{1}{2} \left(\fr12 \z _1{}^2 {\z}_1+2 \dot{\z} _2\right),\quad
 \Phi _3=\fr1{96}\Big[ -\fr{11}4 \z _1{}^4 -8 \z _2^2 -12 {\z}_1 \dot{\z} _2\Big],\quad
\dot{F}_3=-\fr12\z_1\dot{\z}_2+\fr12 \z_1\ddot{\z}_2   \la{csol}   \la{cmodes} \nn\\ 
\eea
Note that the conditions
\bea
\z _0=0,\quad  \Phi_0=0, \quad F_0=1
\eea
reflect the Dirichlet boundary conditions.

\section{`Violation' of Dirichlet boundary condition}

To obtain the quantum-corrected field equations, one should first compute the relevant one-loop diagrams and obtain the one-loop 1PI action. One can then obtain the quantum corrected field equations from the 1PI action in the standard manner.

We focus on the divergent parts associated with the gamma function $\G(\ve)$ where $\ve$ denotes $\ve=2 -\fr{D}{2}$. The gamma function $\G(\ve)$ can be expanded 
\bea
\G(\ve)=\fr1{\ve}-\g+\cdots \la{ge}
\eea
where $\g$ denotes the Euler constant.
Although renormalization removes divergences, the remaining finite terms remain undermined until one imposes further conditions: they get fixed by a renormalization scheme.
Our focus will be these finite terms. The minimal subtraction (MS) scheme removes only the $\fr1{\ve}$ term. The modified MS scheme $\overline{\mbox{MS}}$ is popular in dimensional regularization and it removes the $\fr1{\ve}$ term with additional finite parts (such as $\g$ and $\log 4\pi$). More generally one can employ, as we will, the generalized $\overline{\mbox{MS}}$ and keep the finite parts unspecified.

\vspace{.2in}

\ni Let us shift the metric and scalar  according to \cite{Park:2015ota}
\bea
g_{\m\n}\equiv  h_{\m\n}+\tilde{g}_{{}_B\m\n}\quad \mbox{where}\quad \tilde{g}_{{}_B\m\n}\equiv \vf_{{}_B\m\n}+g_{0\m\n} \la{gshift}
\eea
and
\bea
\z\rightarrow \z_B+{\z} 
\eea
where $g_{0\m\n}$ denotes the classical solution (which will be taken to be a flat spacetime for the divergence analysis) and the fields with the subscript `$B$' denote the background fields. 
As far as the graviton external legs are concerned we limit the analysis to the diagrams with up to (and including) two legs.
The relevant diagrams are as follows:

\subsubsection*{pure gravity sector}

There are two diagrams with two external graviton legs at one-loop:
\bea
&&
\begin{tikzpicture}[line width=1 pt, scale=1]
\draw[vector] (0:-1)--(0,0);
\draw[vector] (.52,0) circle (.49cm);	



\begin{scope}[shift={(1,0)}]
\draw[vector] (0:1)--(0,0);
\node at (0:1.4) {$\vf_B$};
\node at (0:-2.4) {$\vf_B$};
\node at (70:-1.4) {(a) graviton loop };
\end{scope}
\end{tikzpicture} 
\hspace{.2in}
\begin{tikzpicture}[line width=1 pt, scale=1]
\draw[vector] (0:-1)--(0,0);
\draw[dashed] (.52,0) circle (.49cm);	



\begin{scope}[shift={(1,0)}]
\draw[vector] (0:1)--(0,0);
\node at (0:1.4) {$\vf_B$};
\node at (0:-2.4) {$\vf_B$};
\node at (70:-1.4) {(b) ghost loop};
\end{scope}
\end{tikzpicture} \nn\\
&&\hspace{1in}
\mbox{Figure 1   gravity sector}\nn
\eea

\subsubsection*{scalar-involving sectors}

There are three diagrams with two external graviton legs, one of which is
\bea
&&\begin{tikzpicture}[line width=1 pt, scale=1]
\draw[vector] (0:-1)--(0,0);
\draw (.52,0) circle (.49cm);	

\node at (-22:1.3) {$\pa\z$};
\node at (20:1.3) {$\pa\z$};

\node at (-80:.5) {$\z$};
\node at (80:.6) {$\z$};

\begin{scope}[shift={(1,0)}]
\draw[vector] (0:1)--(0,0);
\node at (0:1.4) {$\vf_B$};
\node at (0:-2.4) {$\vf_B$};
\node at (70:-1.4) {Figure 2: scalar loop };
\end{scope}
\end{tikzpicture} \nn
\eea
The other two diagrams are the ones with $(\z,\pa\z)$ lines are replaced by the $(\z,\z)$ lines and $(\pa\z,\pa\z)$ lines, respectively. 
The diagrams with both the scalar and graviton lines are also relevant:
\bea
&& \hspace{.5in}
\begin{tikzpicture}[line width=1 pt, scale=1]
\draw (-140:1)--(0,0);
\draw (140:1)--(0,0);
\draw[vector] (.52,0) circle (.49cm);	

\node at (-140:1.4) {$\pa\z_B$};
\node at (140:1.4) {$\pa\z_B$};

\begin{scope}[shift={(1,0)}]
\draw[vector] (0:1)--(0,0);
\node at (0:1.4) {$\gt_B$};
\node at (70:-1.2) {(a)};	
\end{scope}
\end{tikzpicture} 
\hspace{.4in}
\begin{tikzpicture}[line width=1 pt, scale=1]
\draw (-140:1)--(0,0);
\draw (140:1)--(0,0);
\draw[vector] (.52,0) circle (.49cm);	

\node at (-140:1.2) {$\z_B$};
\node at (140:1.2) {$\z_B$};

\begin{scope}[shift={(1,0)}]
\draw[vector] (0:1)--(0,0);
\node at (0:1.3) {$\gt_B$};
\node at (70:-1.2) {(b)};	
\end{scope}
\end{tikzpicture} \nn\\
&&\hspace{.2in} \mbox{Figure 3: diagrams with both the scalar and graviton lines}  \nn
\eea
The detailed computation of these diagrams will be given in \cite{Park:2016vam}. With the divergence analysis one can show that the one-loop 1PI action takes\footnote{As will be shown in \cite{Park:2016vam}, the above diagram can be absorbed by the following field redefinition,
	\bea
	g_{\m\n}&\ra&  g_{\m\n}+\k^2\Big[l_0g_{\m\n}+l_1  g_{\m\n}R+l_2 R_{\m\n}+l_3 g_{\m\n}(\pa\z)^2+l_4 \pa_\m\z \pa_\n\z+l_5g_{\m\n}\z^2\nn\\
&&+\k^2\Big(l_6R\pa_\m\z \pa_\n \z+l_7R_{\m\n}(\pa\z)^2+l_8R_{\m\n}\z^2+l_9g_{\m\n}R(\pa\z)^2 \nn\\
&& \;\;\;\hspace{.5in}+  l_{10} g_{\m\n}R^{\a\b}  \pa_\a \z\pa_\b \z+l_{11}R^{\a}{}_{\m\n}{}^{\b}  \pa_\a \z\pa_\b \z+l_{12} g_{\m\n}R\z^2\Big)+\cdots \Big]   \la{msg}
	\eea
with
\bea
l_3=l_4=l_5=0,\quad l_2=-e_3,\quad l_1=(\fr12e_3+e_2)
\eea
to the leading order in $\k^2$and thereby the action can be made one-loop renormalizable. Other constraints will appear as one considers higher orders of $\k^2$. The violation of the Dirichlet boundary condition is also hinted at by this redefinition: if one substitutes the classical solution \rf{sersol} with \rf{csol} into the right-hand side of \rf{msg}, the time-dependent terms without any $z$-factor appear. More details will be presented in \cite{Park:2016vam}.}: 
\bea
\G &=&\fr1{\k^2}\int d^4 x \sqrt{-g}\Big[R+\frac{6}{L^2}+e_1\k^4 R\z^2+e_2 \k^2R^2+e_3\k^2 R_{\m\n}R^{\m\n}+\cdots\Big] \nn\\
&& -\int d^4 x \sqrt{-g}\Big[\fr12(\pa_\mu \z)^2 +\fr12m^2\z^2\Big]
\eea
The coefficients $e_1,e_2,e_3$ are finite and depend on the renormalization scheme to be employed. Note that the terms $\z^2 R,R^2$ and $R_{\m\n}R^{\m\n}$ have been inserted as the counterterms unlike the higher derivative approach \cite{Stelle:1976gc} or effective field theory approach \cite{Donoghue:1994dn} where these terms are present already at the classical level. (In other words, the coefficients $e$'s contain $\hbar$ and they are the differences between the divergences and counterterms taken according to one's renormalization scheme.)
The quantum-corrected metric and scalar field equations are
\bea
&& R_{\m\n}-\fr12Rg_{\m\n}-\fr3{L^2}g_{\m\n}-\fr12g_{\m\n}{ \k^2} \Big(-\fr12 (\pa_\mu \z)^2 -\fr12m^2 \z^2 +e_1 { \k^2}R\z^2 -2 e_1{ \k^2}\nabla^2\z^2  \nn\\
&&\hspace{2.5in} +e_2R^2-4e_2\nabla^2 R  +e_3 R_{\a\b}R^{\a\b}-e_3 \nabla^2 R \Big)  \nn\\
&&\hspace{-.1in}+ { \k^2}\Big(-\fr12\pa_\m \z \pa_\n\z+e_1{ \k^2}R_{\m\n}\z^2-e_1 { \k^2}\nabla_\m\nabla_\n \z^2+2e_2 RR_{\m\n}  -2e_2 \nabla_\m \nabla_\n R\nn\\
&& \hspace{1.5in} -2e_3 R_{\k_1\m\n\k_2} R^{\k_1\k_2}-e_3\nabla_\m\nabla_\n R+e_3\nabla^2 R_{\m\n} \Big) =0 
\la{qceom}
\eea
\bea
\nabla^2\z-m^2\z+2e_1\k^2 R\z=0  
\eea
Strictly speaking, one must deal with the boundary terms analogous to the Gibbons-Hawking boundary term. In other words, the generalized Gibbons-Hawking boundary terms will be needed. As a matter of fact, such term have been analyzed in the literature \cite{Deruelle:2009zk,Teimouri:2016ulk}. We will have more on this below.

Let us write down the ansatz as a sum of the classical part plus the corresponding quantum corrections:
\bea
F(t,z)&=& 1 +F_1(t) z+ F_2(t)z^2+F_3(t)z^3 + ...\nonumber\\
&+& (F_0^h(t) +F_1^h(t) z+ F_2^h(t)z^2+F_3^h(t)z^3 + ...) \nn\\
\Phi(t,z)&=&\frac{1}{z}+\Phi_0(t) +\Phi_1(t) z+ \Phi_2(t)z^2+\Phi_3(t)z^3 + ...\nonumber\\
&+& (\fr{\Phi_{-1}^h(t)}{z}+\Phi_0^h(t) +\Phi_1^h(t) z+ \Phi_2^h(t)z^2+\Phi_3^h(t)z^3 + ...)\nn\\
\z(t,z)&=&\z_0(t) +\z_1(t) z+ \z_2(t)z^2+\z_3(t)z^3 + ...\nonumber\\
&+&(\z_0^h(t) +\z_1^h(t) z+ \z_2^h(t)z^2+\z_3^h(t)z^3 + ...)
\eea
As in \cite{Murata:2010dx} one may first explore the case of $\Phi_0(t)=0$ with the similar condition for its quantum counterpart:
\bea
\Phi_0(t)=0\quad,\quad \Phi_0^h(t)=0 \la{Diricon}
\eea
Although it was stated in \cite{Murata:2010dx} that $\Phi_0(t)=0$ is a gauge condition, we view it as a choice consistent with the Dirichlet boundary condition. The classical part of the analysis reproduces \rf{csol}. Interestingly, however, the quantum dynamics constrains the $\z_1,\z_2$ so that
it turns out that\footnote{This feature is also true for the complex scalar case.} 
\bea
\z_1(t)=0\quad ,\quad \z_2(t)=0 \la{psiz}
\eea  
With these, the classical part of the metric becomes just that of AdS spacetime.\footnote{This is intriguing: the classical black hole solution is not sustained. We will have more in the conclusion.}
For the quantum modes one gets\footnote{These and some of the results below were obtained with help of the Mathematica package `diffgeo.m.'} 
\bea
\z_0^h=0,\quad F_0^h=0,\quad
\Phi_{1}^h=0,\quad F_1^h=-2\dot{\Phi}_{-1}^h,\quad \Phi_2^h=0,\quad
F_2^h =0,\quad \z_3^h=\dot{\z}_2^h \nn\\
\eea
The presence of the modes $\Phi_{-1}^h$ implies that the quantum-corrected solution no longer satisfies the Dirichlet condition. (As we will analyze below it can neither be interpreted as a Neumann-type boundary condition.) For one thing, the presence of such modes implies that there may be solutions with a variety of boundary conditions different from the Dirichlet. It also reflects the nontrivial dynamics on the boundary and the information therein stored since the bulk modes depend on it.

Let us explore the case in which we do not impose \rf{Diricon}:
\bea
\Phi_0(t)\neq 0\quad,\quad \Phi_0^h(t)\neq 0 \la{nDiricon}
\eea
Even with these, \rf{psiz} is not avoided, and the following relations result: 
\bea
&&F_0=1,\quad \z_0=0,\quad \Phi _1 { =0},\quad F_1=2\Phi_0 \nn\\
&&\Phi _2=0,\quad F_2 { = 
\frac{1}{4} \left(4  \Phi _0{}^2-8 \dot{\Phi} _0 \right) }\nn\\
&& \dot{F}_3=0,\quad \z_3=0,\quad \Phi_3=0  ,\quad F_4=-F_3\Phi_0
\eea
and
\bea
&&\hspace{-.3in}\z_0^h=0,\quad F_0^h=0,\quad
\Phi_{1}^h=0,\quad F_1^h=2 \left(-\Phi _0(t) {\Phi}_{-1}^h- \dot{\Phi}_{-1}^{h}+{\Phi}_0^h\right),\nn\\
&&\Phi_2^h=0, \quad F_2^h=-2 \Big(-\Phi _{-1}^h \dot{\Phi} _0+\Phi _0^2 \Phi_{-1}^h-\Phi _0 \Phi_0^h+\dot{\Phi}_0^{h}\Big),
\nn\\
&&\hspace{-.5in}\quad \z_3^h=\Phi _0 \dot{\zeta}_1^{h}+\zeta_1^h \dot{\Phi} _0-\zeta_1^h \Phi _0^2-2
\zeta_2^h \Phi _0+\dot{\zeta}_2^{h},\quad { \dot{F}_3^h}=-3F_3 \dot{\Phi}_{-1}^{h},\quad
 \Phi_3^h=0\nn\\
\eea
Note that the modes $ \z_1^h$ and $ \z_2^h$ are not constrained.

The fact that the presence of the terms such as $\Phi_{-1}^h(t)$ cannot be interpreted as a Neumann boundary condition can be seen as follows.\footnote{To be complete, all of the surface terms arising from the quantum correction terms in \rf{qceom} must be examined as well. Those terms would be cancelled by the generalized boundary terms in the extension of the standard practice. As previously pointed out, such analyses have been conducted and can be found in \cite{Deruelle:2009zk,Teimouri:2016ulk}. In this work we limit the analysis to the Einstein-Hilbert term for technical simplicity; we do not expect that the generalized Gibbons-Hawking terms will change the qualitative conclusion of the present analysis. The fact that more and more boundary terms would have to be subtracted out in order to maintain the Dirichlet boundary condition and that there may be a variety of other possible boundary conditions seems to point towards the full boundary dynamics on its own, which is in the same spirit, e.g., as with AdS/CFT and \cite{Freidel:2016bxd,Donnelly:2016auv}. } Suppose one does not add the Gibbons-Hawking term. In this case one should check whether or not the boundary terms \rf{veh} vanish by a (non-Dirichlet) boundary condition.
The second term in \rf{veh} is the trace piece and therefore can be set apart \cite{Park:2015xoa}; let us focus on
\bea
\int d^4x \sqrt{-g}\; \nabla_{\a}  \nabla_{\b}\,\d g^{\a\b} \la{bdtr}
\eea
Since the boundary is specified by a value of the $z$-coordinate, the expression \rf{bdtr} yields
\bea
\int d^3y \sqrt{-g}\;   \nabla_{\b}\,\d g^{z\b} 
\eea
As will be presented in detail in \cite{Park:2016vam} the presence of the factor $\sqrt{-g}\sim \fr1{z^4}$ overpowers the other factors and the expression above cannot be made to vanish by itself (say, by a would-be Neumann-type boundary condition). In other words, it should be removed by the Gibbons-Hawking term. 
However, the Dirichlet boundary condition is not, as shown above, respected by the quantum corrections. The boundary dynamics must be coupled to the bulk in a nontrivial way and has more contents than can be handled simply by a Neumann-type boundary condition.

\section{Conclusion}

In this work we have studied quantum gravitational effects on the boundary and boundary conditions and shown that the quantum effects 'violate' the Dirichlet boundary condition imposed at the classical level black hole solution. With the observation that the classical boundary term cannot be removed by, say, a Neumann-type boundary condition, this has led us to conclude that 
the boundary has the full dynamics of its own that are coupled to the bulk, which is consistent with the AdS/CFT type dualities. The quantum effects in the bulk leave their footprints on the boundary: with the quantum effects the boundary terms become relevant and store the information on the bulk. Narrowing down to the Dirichlet boundary condition and excluding all the other possible boundary conditions may be behind the information loss puzzle, which seems in line with the view expressed in \cite{Freidel:2016bxd,Donnelly:2016auv}.
The result of the present work support the anticipation in \cite{Park:2014mba} that the information storage mechanism should be of the teleological character because the time-dependence of the bulk and boundary modes are determined simultaneously.  
A field redefinition similar to that of the present work is also needed in the quantization scheme recently proposed in \cite{Park:2016zgt} (and refs therein), and will have the same interpretation as in the present work. 

One of the future directions is to understand the intriguing feature of our results: the classical black hole solution is not sustained at the quantum level in the manner naively expected. Presumably it reflects the instability of a black hole against the Hawking radiation.

\vspace{.4in}

\ni {\bf Acknowledgments}

\ni I thank Daniele Pranzetti for sharing his expertise.

\newpage

\end{document}